\documentclass[twocolumn,twocolappendix]{openjournal}
\renewcommand{\baselinestretch}{1.075}
\shortauthors{Maloney et al.}
\graphicspath{{./}{figures/}}

\usepackage{macros}
\usepackage[dvipsnames]{xcolor}
\usepackage{amsmath}

\newcommand{\pibarunit}[0]{\mu as ~ Mpc ~ yr^{-1}}

\begin{document}
\title{Constraining Extragalactic Proper Motion with Gaia Astrometry}

\author{Sean Maloney}
\affiliation{Physics and Astronomy Department, University of Pittsburgh,
Pittsburgh, PA 15260, USA}

\author{Tianqing Zhang}
\affiliation{Physics and Astronomy Department, University of Pittsburgh,
Pittsburgh, PA 15260, USA}

\author{Rupert Croft}
\affiliation{The McWilliams Center for Cosmology \& Astrophysics, \\ Department of Physics, Carnegie Mellon University, Pittsburgh, PA 15213, USA}

\author{Konstantin Malanchev}
\affiliation{The McWilliams Center for Cosmology \& Astrophysics, \\ Department of Physics, Carnegie Mellon University, Pittsburgh, PA 15213, USA}

\begin{abstract}
    The Solar System's secular motion with respect to the cosmic microwave background (CMB) rest frame is inferred from the CMB dipole and should induce a tiny, coherent apparent drift in the positions of nearby galaxies, referred to as the extragalactic proper motion. We test the feasibility of a purely geometric measurement of this effect by combining Gaia DR2 and DR3 astrometry with low-redshift spectroscopic galaxy catalogs to build a large, full-sky sample of $67,173$ galaxies.
    Although we do not obtain a statistically significant detection of the expected dipole signal, we place the tightest constraint to date on the extragalactic proper motion $\bar{\pi}$. Using galaxies with comoving distance $D>5 {\, \rm Mpc}$, we also place the tightest constraints on cosmic extragalactic proper motion $\bar{\pi}_{\rm cosmic}$, with uncertainty $\sim 10\times$ the measured CMB dipole value. Our $1\sigma$ uncertainty on the near field extragalactic proper motion $\bar{\pi}_{\rm nf}$ is approximately $\sim 1.3\times$ the expected CMB measurement, demonstrating that Gaia astrometry is approaching the sensitivity required for a direct detection of near field Hubble constant in future releases.
\end{abstract}

\section{Introduction}


The motion of the Solar System with respect to the cosmic microwave background (CMB) rest frame is one of the most precisely measured quantities in cosmology. It is observed as the dominant dipole anisotropy in the CMB temperature map and corresponds to a velocity of approximately $370~\mathrm{km\,s^{-1}}$ toward $(\alpha, \delta) \approx (168^\circ, -7^\circ)$ \citep{Lineweaver1997}. This motion is a combination of the effects of the Earth’s orbit, the Sun’s motion within the Milky Way, and the peculiar velocity of the Milky Way itself.

While the CMB dipole provides a measurement of this motion in the radiation rest frame, it also predicts a purely geometric observable from the nearby galaxies. As the observer moves through space, the apparent positions of extragalactic objects must slowly drift on the sky. This effect, often referred to as \emph{extragalactic proper motion}, is entirely kinematic and geometric and barely relies on any astrophysical modeling of the sources themselves. From the CMB dipole, the expected signal is extremely small, approximately
\begin{equation}
\bar{\pi} \approx 77.8~\mu\mathrm{as}\,\mathrm{Mpc}\,\mathrm{yr}^{-1},
\end{equation}
for objects located $90^\circ$ from the dipole \citep{Croft2021}, but it has a specific angular and redshift dependence that makes it, in principle, measurable with nearby galaxies.

Several studies have previously explored the possibility of detecting this effect. Early theoretical work established the expected signal and its geometric dependence\citep{Kardashev1973, Ding2009}, and more recent studies (e.g., \citealt{Paine2020}) attempted observational constraints using available astrometry data in Gaia DR2 \citep{2018GaiaDR2Paper}. However, these studies were limited by small sample sizes and insufficient astrometric precision. The resulting constraints were far from the expected signal amplitude, with upper limits on the Hubble parameter inferred from cosmic proper motion exceeding the CMB-based value by factors of $\sim 40$ \citep{Paine2020}.

The advent of the \emph{Gaia} mission \citep{2016GaiaMission} fundamentally changes this landscape. Gaia provides micro-arcsecond astrometry for billions of objects across the sky, including millions of galaxy candidates \citep{gaia_extragalactic_dr3}. Although Gaia was designed primarily for stellar astrometry, its repeated all-sky measurements and long temporal baseline make it uniquely suited for detecting tiny coherent shifts in the positions of extragalactic sources. Crucially, Gaia Data Release 2 (DR2; \citealt{2018GaiaDR2Paper}) and Data Release 3 (DR3; \citealt{gaia_extragalactic_dr3}) provide two well-separated epochs of astrometry, enabling a direct measurement of positional drift over a baseline of roughly 1 year.

In this work, we investigate whether the extragalactic proper motion can be detected directly from Gaia using the astrometry of nearby galaxies. We construct a large, full sky catalog by cross-matching Gaia DR2 and DR3 with spectroscopic galaxy redshift surveys, specifically the 2MRS Redshift Survey \cite[2MRS][]{Huchra2012} and the 6dF Galaxy Survey \cite[6DFGS][]{Jones2009}. This cross-match provides both high-precision astrometry and reliable redshift estimates, allowing us to exploit the predicted redshift dependence of the cosmic proper motion signal. In this work, we are not attempting to make a Hubble parameter measurement; we thus use redshift as an estimate of the distance by assuming a Hubble parameter of $70 \, {\rm km\, /s / Mpc}$.

Instead of relying on proper motion measured by the Gaia data release, we cross-match Gaia DR2 and DR3 and use the difference in the centroid between the releases as the proper motion. 
As a result, our approach benefits from a dramatic increase in sample size and sky coverage. Our final catalog contains 67,137 galaxies, nearly three orders of magnitude larger than those used in earlier attempts \citep{Paine2020}. This enables a substantial reduction in statistical uncertainty. We carefully model the astrophysical and observational uncertainty, such as velocity correlations and the covariance matrix of the coherent motion of the physically associated galaxies in the sample.

We fit a geometric cosmic proper motion model to the observed change in angular separation of each galaxy relative to the CMB dipole direction between GAIA DR2 and DR3. Although our results do not yet constitute a definitive detection of the signal, we achieve a significant improvement in sensitivity and demonstrate that Gaia astrometry is approaching the regime where a detection becomes feasible. Our work provides both an empirical validation of the cosmic proper motion framework and a roadmap for how future Gaia releases and upcoming facilities such as the Nancy Grace Roman Space Telescope and the Vera C. Rubin Observatory could enable a clear measurement of this effect.

\begin{figure*}
    \includegraphics[width=2.0\columnwidth]{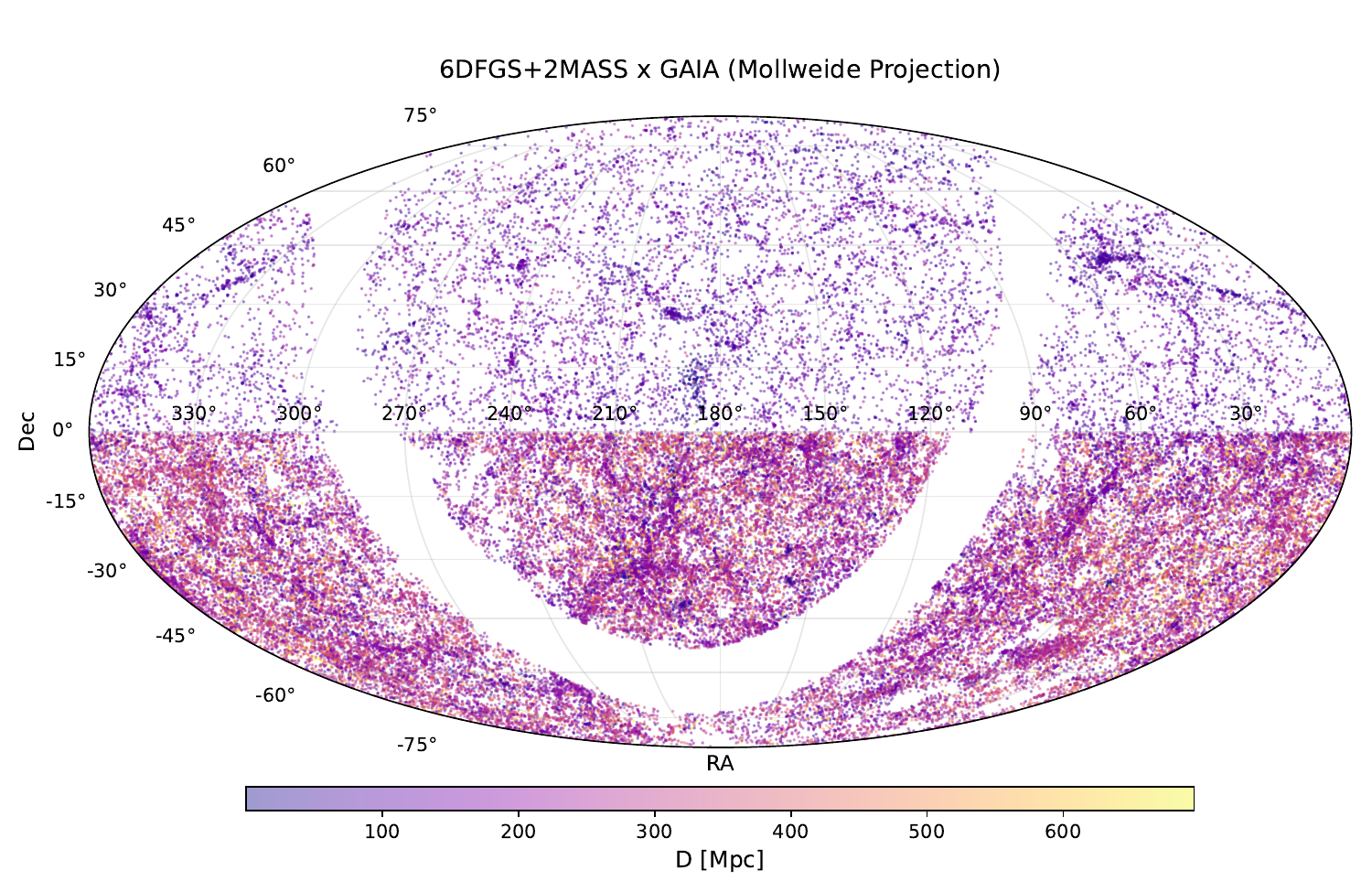}
    \caption{\label{fig:scatter} Sky distribution in equatorial coordinates (RA, Dec) using a Mollweide projection for the cross-matched sample between 6dFGS+2MRS and Gaia, color coded by their distance. The galactic plane has been masked to avoid high-extinction regions and star contamination. The abundance of southern hemisphere objects reflects the sky coverage of the 6dFGS survey.}
\end{figure*}

This paper is organized as follows. In Section~\ref{sec:data:0} we describe the construction of the extragalactic proper motion catalog from the Gaia extragalactic catalog and spectroscopic surveys. Section~\ref{sec:method:0} details the cross-matching procedure, the cosmic proper motion model, and our treatment of measurement and systematic uncertainties. Section~\ref{sec:results:0} presents our key results on the near-field and cosmic extragalactic proper motion, and consistency tests on our measurement. We conclude in Section~\ref{sec:conclusion:0} with a discussion of the implications for future astrometric surveys and prospects for a definitive detection of cosmic proper motion.

\section{Extragalactic Proper Motion Catalog}
\label{sec:data:0}

    Since the extragalactic proper motion decays inversely with distance, only the nearest galaxies bear its signal. 
    With that in mind, data for this study are carefully selected to have high accuracy and contain only extragalactic sources of low redshift. Due to the nature of the apparent motion of a galaxy being very small on short timescales (timescale discussed in section \ref{sec:time-scale}), our signal is highly sensitive and requires that our astrometry and distances be highly accurate. Because of these requirements, we opted to create the extragalactic proper motion catalog from a variety of surveys. The astrometry in our catalog is provided by Gaia \citep{2016GaiaMission} data release 3 \& 2, while distance and redshift data is provided by the Two Micron All-Sky Redshift Survey \citep[2MRS,][]{Huchra2012} and 6-degree Field Galaxy Survey \citep[6DFGS,][]{Jones2009}. Gaia is an all-sky survey with a focus on mapping the dynamic Milky Way, so it contains mostly stellar objects, which we exclude in our extragalactic catalogs. While it is challenging to effectively separate extended objects from stellar objects, overcoming contamination in this work from stellar objects can be achieved as we require our Gaia data to be cross-matched with the 2MRS and 6DFGS redshift catalog, which are both galaxy surveys, thus any non-galaxy objects are dropped in the cross-match. We combined 2MRS and 6DFGS and cross-matched with the entirety of Gaia DR3 and DR2, producing a catalog of 67,137 galaxies. We discuss 2MRS and 6DFGS in \ref{sec:2MRS}, and Gaia in \ref{sec:Gaia}. We further discuss cross-matching and catalog construction in \ref{sec:lsdb}.
     
    \subsection{2MRS and 6dFGS}
        \label{sec:2MRS}
        
        To detect the cosmic proper motion dipole, we require a local, wide-area galaxy redshift catalog that provides robust three-dimensional positions of nearby galaxies. Two complementary datasets fulfill this criterion: the Two Micron All-Sky Survey \citep[2MRS,][]{Huchra2012} Redshift Survey and the 6-degree Field Galaxy Survey \citep[6DFGS,][]{Jones2009}. We combine these to construct a catalog with broad sky coverage and reliable redshift information at low redshifts ($z \lesssim 0.1$).
        
        The 2MRS Redshift Survey covers the entire sky in the near-infrared and provides spectroscopic redshifts for galaxies selected from the 2MRS Extended Source Catalog. We use a subset of the 2MRS catalog with 43,507 galaxies and restrict our selection to the northern sky (Declination $> 0$) to avoid overlapping coverage with 6dFGS and to ensure complementarity.
        
        The 6dFGS provides redshift measurements over much of the southern sky. We use the publicly released 6dFGS Data Release 3, which contains redshifts for 124,647 galaxies. For each galaxy in the 6dFGS catalog, we compute the comoving distance using the fiducial cosmology adopted in this analysis. The relevant subset of the catalog includes right ascension, declination, redshift, and comoving distance.
        
        The two catalogs are pre-processed as follows:
        \begin{itemize}
        \item The 2MRS catalog is read from a FITS file containing 43,507 galaxies. We extract RA, Dec, redshift, and comoving distance columns, and restrict to galaxies with Declination $> 0$.
        \item The 6dFGS DR3 catalog is read from a plain-text table. After computing comoving distances from the redshifts, we extract the same four columns: RA, Dec, $z$, and $D_{\rm com}$. We restrict galaxies with Declination $< 0$ for 6dFGS.
        \item The two catalogs are concatenated to form a single dataset of 130,087 galaxies.
        \end{itemize}
        
        The final combined catalog is saved for subsequent cross-match with the GAIA dataset. This merged sample offers excellent sky coverage and redshift completeness for mapping the nearby large-scale structure and inferring the anisotropic cosmic proper motion signal.

    \subsection{Gaia}
        \label{sec:Gaia}
        
        Gaia is a scanning space astrometry observatory that allows the observatory to self-correct its astrometric solution as it scans the sky. This provides Gaia with highly accurate astrometry, necessary for our study, and thus acts as our primary source for position data. Gaia's data is also available to the public, so we were able to begin looking at test pipelines for the cosmic proper motion detection with Gaia instantly. We used Gaia's two latest data releases, DR2 and DR3, which, when cross-matched, contained 67,137 galaxies. The Gaia data was further cross-matched with 2MRS and 6DFGS to provide redshift data for our catalog. 
        
        \begin{figure}
            \includegraphics[width=1.0\columnwidth]{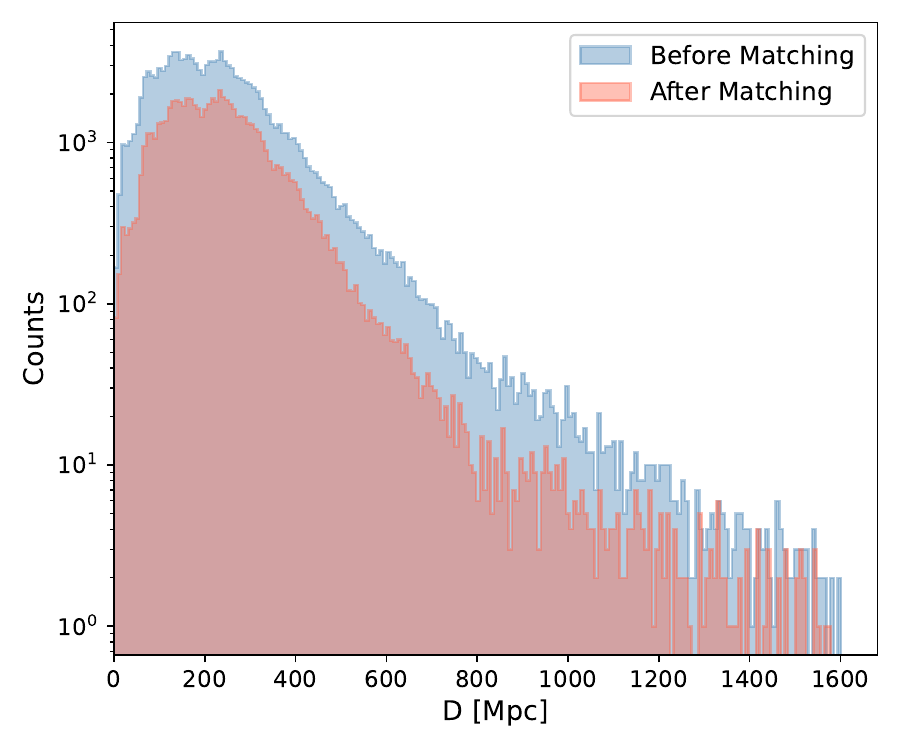}
            \caption{\label{fig:nz} Histogram of distances for galaxies in the 6dFGS+2MRS catalog (blue) and the subset matched with Gaia objects (red). The matching with Gaia reduces the total number of galaxies, particularly at higher distances, but preserves the overall shape of the distribution. Counts are shown on a logarithmic scale.}
        \end{figure}
        
        \subsubsection{Gaia DR3 and DR2}
            Gaia is our main catalog for astrometry in the extragalactic proper motion catalog. We use astrometry Gaia data release 3 \citep[DR3,][]{2023GaiaDR3Paper} and data release 2 \citep[DR2,][]{2018GaiaDR2Paper} as our final and initial positions respectively.
            We use \texttt{LSDB}\footnote{\url{https://github.com/astronomy-commons/lsdb}} \cite{Caplar2025} to cross-match the Gaia DR2 and Gaia DR3 object catalog. 
            After cross-matching, we filter the dataset by the DR3 extragalactic candidate flag provided by \cite{gaia_extragalactic_dr3}. 
            DR3 initially contains 4.8 million possible galaxy candidates.  After our cross-match with DR2, 2MRS, and 6DFGS, our final catalog contained 67,137 objects with a mean redshift of  $z=0.0527$. We show the spatial distribution of the cross-matched catalog in Figure \ref{fig:scatter}, and the redshift distribution of the catalog in Figure \ref{fig:nz}.
    
        \subsubsection{Time scale} \label{sec:time-scale}

            An important component of our analysis is the temporal baseline between the two observations in our extragalactic proper motion catalog. Ideally, one would use the time of observations per-galaxy. As this information is not readily available from the Gaia releases, we attempt to estimate the average time span $\Delta T$ between Gaia DR2 and DR3, and apply $\Delta T$ overall in our modeling. 

            
            Gaia DR3 and DR2 do not directly provide a time of observation for their data, but we are able to find an average time of observation for each data release. DR3 and DR2 are both subsets of the same set of continuous observations with DR2 being a smaller subset based on earlier observations. We can take advantage of Gaia's auxiliary data set \texttt{gaiadr3.commanded\_scan\_law} to get the time at which Gaia observed each part of the sky. Further we are able to see from descriptors about DR3 and DR2 the time period where each dataset was constructed.

            Observations for both data sets began on the 25th of July 2014 with DR2 being completed on the 23rd of May, 2016 and DR3 being completed on the 25th of May, 2017. DR3 spans 34 months of observations and DR2 has 22 months; however, due to our cross-match, any overlap between DR3 and DR2 is negated and observations for DR3 can be approximated as starting at the end of DR2 observations. We find the average time of observation of each data set and find the difference to be approximately $\Delta T = 1.01$ years. 
            

    \subsection{Catalog Cross-Matching with LSDB} \label{sec:lsdb}

        To accurately measure cosmic proper motion, we must identify the same galaxies across multiple surveys with differing resolutions, depths, and astrometric precision. For this purpose, we employ the LSDB (Large Survey Database)\citep{Caplar2025}, a scalable Python-based interface for spatially indexed astronomical catalogs. LSDB handles catalogs in the HATS (Hierarchical Astronomy Tables Structure) format, which organizes survey data into spatially tiled partitions. This structure enables fast positional queries, distributed computing, and high-throughput catalog cross-matching, even for GAIA catalogs which contains billions of objects.
        
        We utilize LSDB to perform two key cross-matching steps:
    
        \begin{enumerate}
        \item \textbf{Gaia DR2–DR3 self-matching:} As described in Section~\ref{sec:Gaia}, we first construct a self-consistent Gaia catalog by cross-matching Gaia Data Release 2 (DR2) with Data Release 3 (DR3). We read the LSDB-formatted DR2 and DR3 catalogs from the HATS directories hosted on the Pittsburgh Supercomputer Center (PSC). We then perform a one-to-one positional cross-match within a $0.02''$ search radius to associate DR2 sources with their DR3 counterparts. 
        \item \textbf{Gaia–2MRS/6dFGS matching:} Next, we cross-match the Gaia DR2–DR3 combined catalog with the 2MRS + 6dFGS spectroscopic sample described in Section~\ref{sec:2MRS}. The combined low-redshift catalog, pre-processed and saved in LSDB-compatible HATS format, is matched against Gaia using a broader $1''$ search radius, retaining only the nearest neighbor per source:
        \end{enumerate}
        
        The final cross-matched dataset includes 67,137 galaxies with proper motions from Gaia and redshift-comoving distances from 2MRS and 6dFGS.
        Figure~\ref{fig:scatter} shows the sky distribution of the final cross-matched catalog between Gaia and the combined 2MRS + 6dFGS sample, plotted in equatorial coordinates using a Mollweide projection. The galactic plane has been masked to mitigate contamination from stellar crowding and extinction, which can compromise the Gaia astrometry measurements. The figure highlights the nearly full-sky coverage of the resulting sample, with an over density in the southern hemisphere due to the footprint of the 6dFGS survey. 
        This catalog forms the basis of our proper motion dipole measurement in subsequent sections.

    \subsection{Data curation} \label{sec:data-curation}

        Cosmic proper motion arises as an apparent motion in extragalactic objects as we move with respect to the CMB dipole. The CMB dipole is defined to be (RA, Dec) = (167.98, -7.22) \cite{Lineweaver1997} and is associated with the direction of travel of the solar system with respect to the cosmic rest frame. In this study, we compare our results to the CMB dipole measurement $\bar\pi = 77.8 \pibarunit$, which we assume to lie along the same axis of the absolute motion of the solar system.
        
        In order to measure the change in position of a galaxy's centroid with respect to the CMB dipole. We define its change in position to be $\Delta\beta = \beta_1 - \beta_2$ where $\beta_1$ is the separation of the DR3 catalog from the CMB dipole and $\beta_2$ is the separation of the DR2 catalog. Because we expect the motion of the Earth to be along the CMB dipole axis, objects close to the CMB dipole will have very little change in radial position, while transverse movement will be large. Because our model only depends on radial movements, we apply a weighting factor to all objects equal to $sin(\beta_1)$ such that objects along the dipole axis will have zero contribution and those $90\degree$ away will have the largest contribution.

\section{Methodology}
\label{sec:method:0}

In this section, we describe the model we use to fit for the extragalactic proper motion, and modeling of its observational and astrophysical systematic uncertainties. 

    \begin{figure*}
    \includegraphics[width=1\columnwidth]{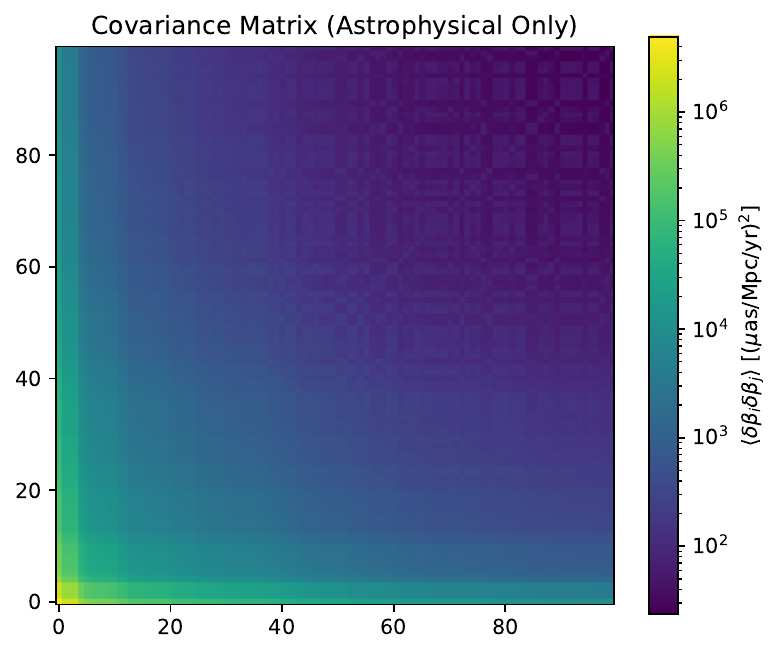}
    \includegraphics[width=1\columnwidth]{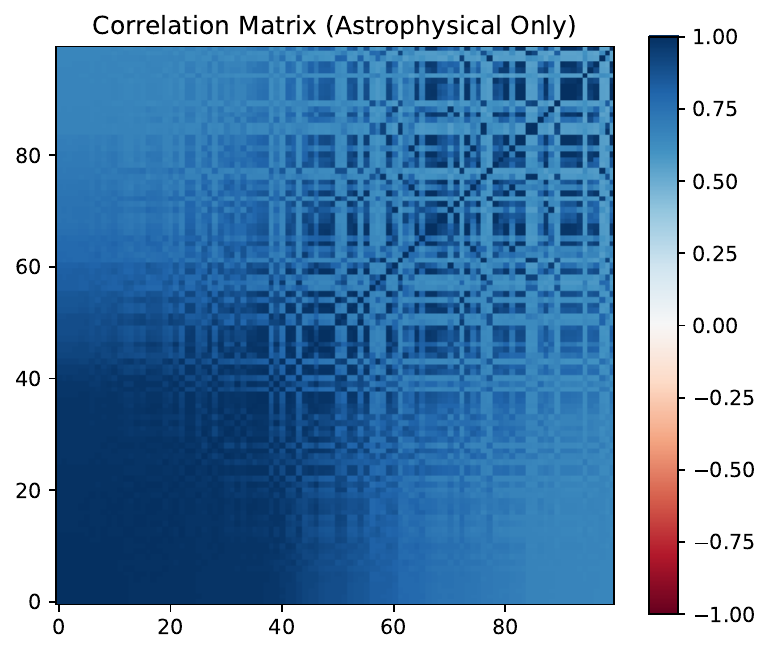} 
    \includegraphics[width=1\columnwidth]{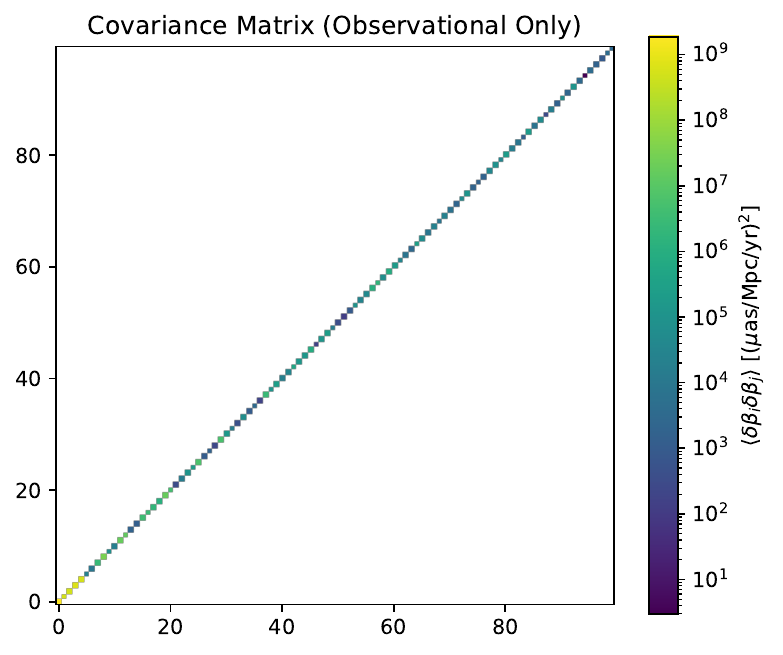} 
    \includegraphics[width=1\columnwidth]{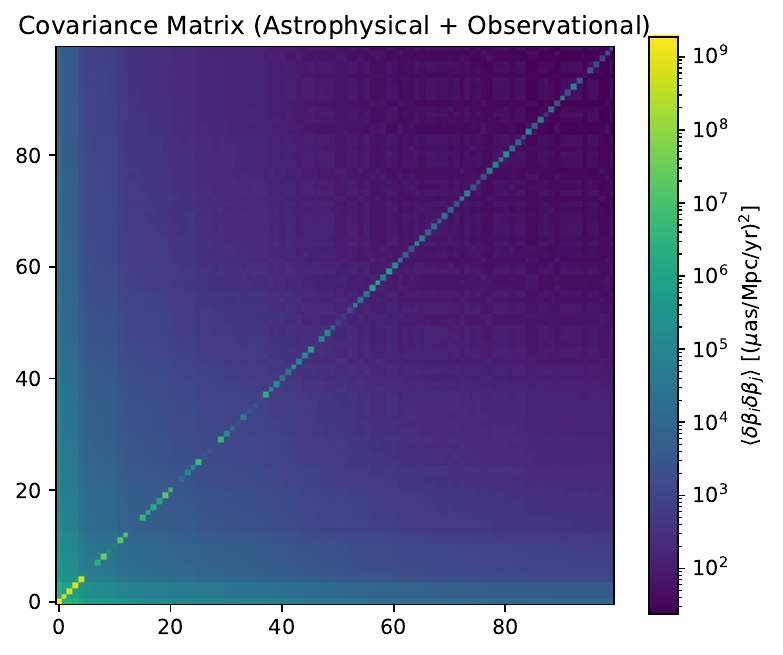} 
    \caption{The peculiar-velocity proper motion covariance matrix $D_{ij}$ (upper left) and its corresponding correlation matrix (upper right), constructed from 100 nearest galaxies DR3 matched catalog, sorted by comoving distance.  The off-diagonal covariance is computed for the 100 nearest galaxy pairs. The observational proper motion uncertainty covariance (lower left) and the combined astrophysical plus observational covariance (lower right) are also shown.}
    \label{fig:cov_corr}
    \end{figure*}

 \subsection{Cosmic proper motion Model}
\label{sec:method:model}

In order to describe the apparent rate of motion of astrophysical objects caused by Earth's motion with respect to the CMB dipole, we construct a simple model which takes in the change in separation from the CMB dipole $\Delta \beta$ (see \ref{sec:data-curation}), the angle between the object and the CMB dipole, the distance of the object to the observer, and the average time between initial and final observations. The model relates a direct measurement of radial motion $\delta\beta$ to a proper-motion parameter $\bar{\pi}$ which describes the proper-motion of an object given its real distance and separation from the CMB dipole
\begin{equation}
    \Delta\beta_i = \bar{\pi} sin(\beta_{1i})\frac{\Delta T}{D_i}.
\end{equation}
To account for potential systematic error that is independent of $D$, we introduce a constant parameter $C$ to the modeling
\begin{equation}
    \frac{\delta\beta_i}{sin(\beta_{1i})} = \bar\pi \frac{T}{D_i} + C
\end{equation}
$D$ is the comoving distance to the astrophysical object and is taken directly from our initial data. $\Delta T \equiv 1.01$ year is the average time between Gaia observations in DR3 and DR2 and is discussed in \ref{sec:time-scale}. $\Delta\beta$ is discussed in the previous section \ref{sec:data-curation} and is the only variable we directly calculate. $\beta_1$ is also discussed in sec. \ref{sec:data-curation}, it refers to the DR3 separation from the CMB dipole. $\bar\pi$ is our dependent variable and is found through a non-linear least squares regression of our model. We fit $\bar\pi$ for N galaxy candidates, where N is the number of cross-matched galaxies.

$\bar\pi$ is the extragalactic proper motion. Relative to the cosmic rest frame, this effect has a measured value of 77.8 $\pibarunit$ for objects $90\degree$ away from the CMB dipole \citep{Croft2021}. In this work, we assume this motion can be measured by using galaxies that are not in the local group, which is defined as $D_i>5 \, {\rm Mpc}$. 

We perform a weighted non-linear least squares regression in order to recover the Earth's motion and perform a $\chi^2$ test on our data. We remove outlier objects for which $\frac{(x_i - \bar x)^2}{\sigma^2}>8$, and inflate the covariance matrix described in Section~\ref{sec:method:uncertainty} so that the reduced $\chi^2/N = 1$.

\subsection{Uncertainty Characterization}
\label{sec:method:uncertainty}

Uncertainty of the extragalactic proper motion has two main components, observational uncertainty and astrophysical uncertainty from galaxy dynamics and their correlation. Astrometric uncertainty is provided by Gaia for RA and Dec in both DR3 and DR2 catalogs. Velocity correlation is a astrophysical uncertainty of the proper motion that arises from the galaxy's Virial and in-fall motion due to their gravitational attraction to each other. We adopt the correlation described in \cite{Lyall_2024}, to model the cross correlation of the galaxy dynamics.  The astrometric uncertainty and velocity correlation must be combined into a covariance matrix, where astrometric error only contributes to the diagonal elements, and the velocity correlation is the variable correlation. In Figure~\ref{fig:cov_corr}, we show the covariance matrix and correlation matrix used in this analysis. The components of the covariance matrix are calculated as follows:
\begin{itemize}
    \item Astrometric error is provided by Gaia for its astrometric measurements. We begin by adding the RA errors for DR3 and DR2 together in quadrature. The same procedure is carried out for DEC. We then again combine these new errors in quadrature to find the magnitude of astrometry uncertainty for each object. The astrometry uncertainty only contributes to the diagonal elements of the covariance matrix, shown in the bottom left panel of the Figure~\ref{fig:cov_corr}.
    \item Velocity correlation is calculated by $V(r)$ provided in \citep{Lyall_2024} between two galaxies and mapping their distance to the fiducial velocity correlation. $V(0)$, which is the variance of the single galaxy is approximated to be $300 \, {\rm km/s}$.  We conduct this calculation for all nearby galaxy-pairs in our catalog by first sorting them via distance and indexing through each object and comparing it to every object in the catalog. The covariance and correlation matrix of the closest 100 galaxies in our extragalactic proper motion catalog is shown in the top panels of Figure~\ref{fig:cov_corr}.
\end{itemize}

The velocity correlation is a two-dimensional covariance matrix where $A_{ij} = 0$ for $i=j$. The square of astrometric error is added along the diagonal to include the random error in our covariance matrix. We are unable to create a full covariance matrix for every object in our catalog as doing so would require creating a covariance matrix of size 67,000 $\times$ 67,000 due to memory limitations. 
To reduce the computational burden, we limit the velocity correlation calculation to the nearest 100th object (D = 9.473 Mpc) and calculate only the astrometry uncertainty for the next 9,900 objects. It was determined that including the velocity correlation for more distant objects had a negligible impact on the outcome of the analysis and did not influence final results.
        
        
        

\section{Results}
\label{sec:results:0}

In this section, we present the least-squares fitting on the extragalactic motion catalog, described in Section~\ref{sec:data:0}, using the model described in Section~\ref{sec:method:model}. We present our key findings -- upper bound of the extragalactic proper motion with and without non-local galaxies, Section~\ref{sec:result:key_finding}. In Section~\ref{sec:result:consistency}, we present the consistency tests with our results. 

  This section will discuss the key findings of the detection, as well as the associated consistency test conclusions. 
    
\subsection{Key Findings}
\label{sec:result:key_finding}

For our main results, we present the results of cosmic proper motion with our fiducial model:
\begin{itemize}
    \item we use a constant parameter in the model;
    \item we calculate the velocity covariance matrix of the galaxies for the closest 10,000 galaxies;
    \item we removed $5\sigma$ outliers in $\frac{\delta\beta_i}{sin(\beta_{1i})}$.
\end{itemize}

\begin{figure}
    \centering
    \includegraphics[width=\linewidth]{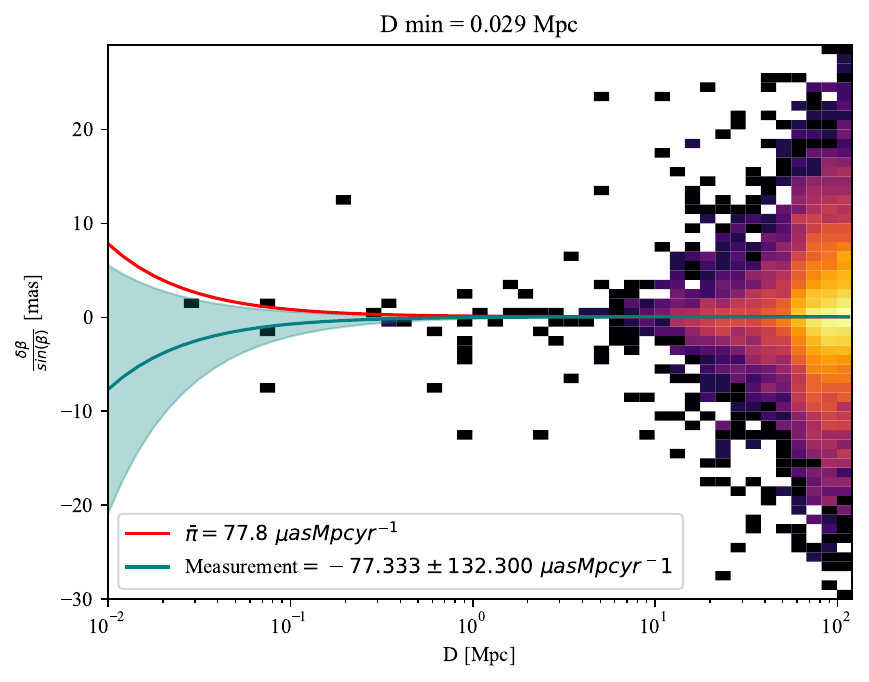}
    \includegraphics[width=\linewidth]{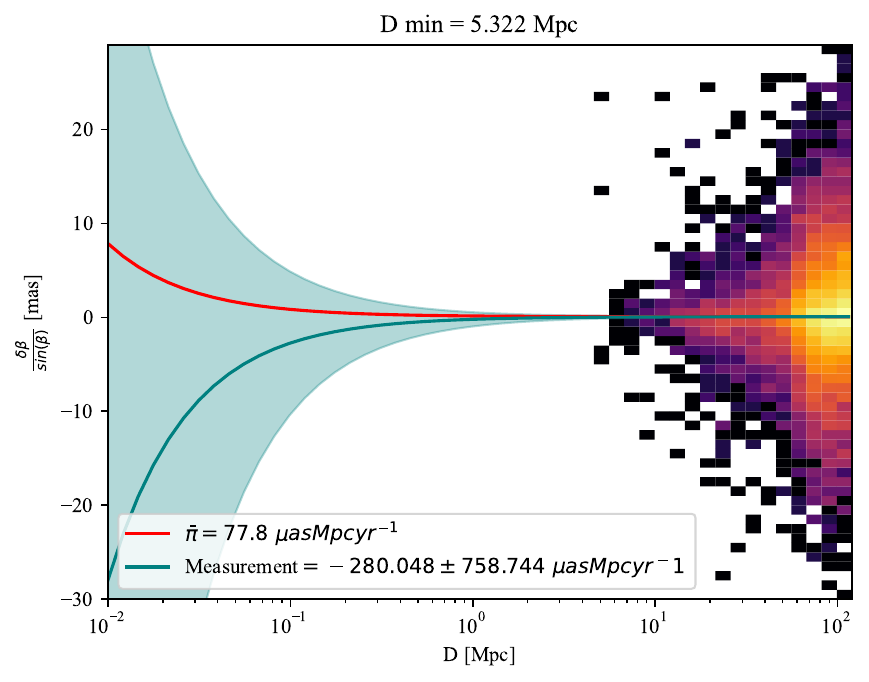}
    \caption{Shows the predicted value of $\bar\pi$ as a function of D and a constant T. The teal \textit{'Measurement'} line shows the calculated value for $\bar\pi$ and the envelope is its error. The red \textit{'Expectation'} line shows the value $\bar\pi = 77.8 \pibarunit$. Note that $\bar\pi$ is shown here in $mas$ rather than $\mu as$ for clarity.}
    \label{fig:near-field-results}
\end{figure}

First, we present near-field extragalactic proper motion constraints, using all galaxies with no cut on minimum distance $D$. 
Our model shows the near-field extragalactic proper motion to be 
\begin{equation}
\bar\pi_{\rm nf} = -77.333 \pm 132.30 ~\pibarunit,
\end{equation}
which is shown in Figure~\ref{fig:near-field-results}. This lies within $1\sigma$ of CMB measurement and marks the tightest $\bar\pi$ constraints with local galaxies to date. While it is still a non-detection of near-field proper motion, we show a significant reduction in the upper bound of this signal. With a dataset that has a longer time baseline to Gaia DR2, e.g., Gaia DR4 or Roman data releases, we are approaching the detection limit of near-field extragalactic proper motion. 

Secondly, we present the constraints on the cosmic extragalactic proper motion, with a distance cut of $D_{\rm min} = 5 \, {\rm Mpc}$. Our model shows a cosmic extragalactic proper motion
\begin{equation}
\bar\pi_{\rm cosmic} = -280.048 \pm 758.744 ~\pibarunit
\end{equation}.
We pick $D_{\rm min} = 5 \, {\rm Mpc}$ as a commonly used scale to exclude galaxies that are in the local group (e.g., applied in \cite{Paine2020}). 
Our $\bar\pi_{\rm cosmic}$ is consistent with the measurement of the CMB, and a null detection. 
The uncertainty on $\bar\pi_{\rm cosmic}$ is $\sim 10\times$ the CMB measurement, which also marks the tightest measurement on this number. 
    
Our large catalog size was able to greatly reduce the upper limit on the cosmic proper motion signal compared to earlier attempts, as well as greatly reduce the statistical error. With future data releases from Gaia as well as high-quality data from upcoming telescopes like NGRST, we predict that a direct measurement of cosmic proper motion will soon be possible.
    
\subsection{Consistency Tests}
\label{sec:result:consistency}

\begin{figure*}
\includegraphics[width=\columnwidth]{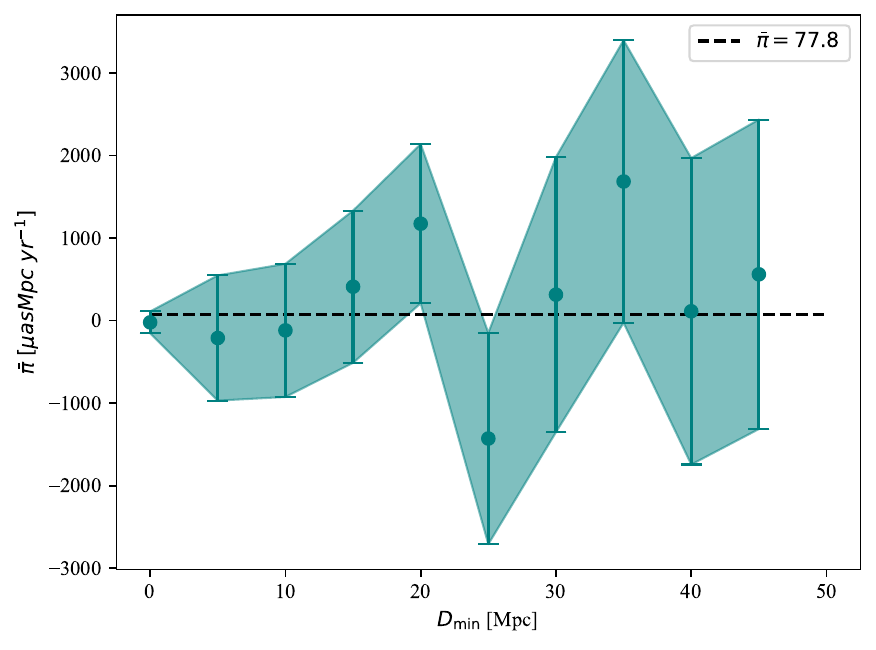}
\includegraphics[width=\columnwidth]{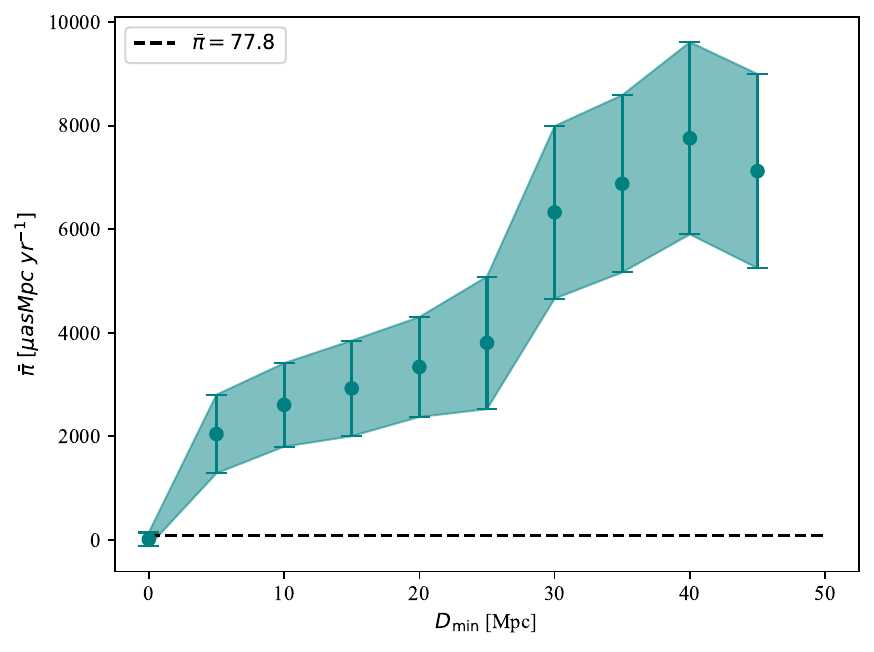}
\caption{Consistency test of various models with different $D_\text{min}$ cuts. The left panel shows models with a constant (our fiducial choice), while the right panel shows those with a constant. We remove galaxies with reduced $chi^2$ of $\frac{\delta\beta}{\text{sin}(\beta)}$ greater than 8. The models that included the constant term show more consistency between different $D_{\rm min}$ values applied, therefore are picked as fiducial analysis choices. }
\label{fig:subim3}
\end{figure*}

Because our results are sensitive to galaxy samples of close distance, we re-ran our model for many different cutoffs of the minimum distance value. In Fig.~\ref{fig:subim3}, we show the model fitting results with $D_{\rm min}$ out to 50 Mpc. The uncertainty on $\bar{\pi}$ increases significantly when removing the closest galaxies, but the results remain consistent with the CMB dipole prediction. 

In Figure~\ref{fig:subim3}, we also test alternative settings of our model. In the right panel, we show that not having a constant term in the model produces significant $D$-dependent $\bar{\pi}$, which is unphysical and suggests that there are systematic errors. We show that with a constant term and outlier removal at $8\sigma$, the key results of $\bar{\pi}$ is independent of the $D_{\rm min}$ cut value, and therefore picked as the fiducial modeling choice.

As a validation to our model fitting technique, we also generated a simulated mock catalog consisting of 10,000 galaxies with a synthetic $\delta \beta$ signal equivalent to $\bar{\pi} = 77.8 \pibarunit$ plus a random error of $5 mas$ on $\delta \beta$. We confirm that we are  able to recover the measurement of $\bar{\pi}$ within statistical fluctuation. This further solidifies our model-fitting process.

\section{Conclusion}
\label{sec:conclusion:0}
Extragalactic proper motion is a geometric alternative detection of the CMB dipole; it is the measurement of the motion of the Earth with respect to the CMB by observing the change in CMB rest frame positions of nearby galaxies. Accurate astronomical solutions are essential for this detection, and as such, we use data from Gaia, which provided us with the large timescale ($\sim1.1$ yrs) and accurate astrometry. However, we needed to supplement this data with the redshift measurement from the 6DFGS and 2MRS galaxy surveys. The combination of datasets enabled this study to conduct the measurement of this $\mu$-arcsecond scale signal.

Although our detection is not statistically significant, we obtain the best detection of the near-field extragalactic proper motion $\bar{\pi}_{\rm nf}$ to date, with a reduction of $\sim 30\times$ in $\bar\pi$ amplitude from previous best detections. Our result of $\bar{\pi}_{\rm nf} = -12.89 \pm 102.29 ~\pibarunit$ is consistent to $1\sigma$ with the expected signal for $\bar\pi$.
Our detection was limited by difficulties in obtaining proper motions for these galaxies. Gaia, as it was designed to study the dynamics of stars, only provides a limited number of proper motions for extended object candidates, due to failure of its 5-parameter proper motion fitting. As a workaround of this issue, we simply cross-matched Gaia DR2 and DR3 galaxy candidates, and use the difference in the sky coordinate between the data releases as the proper motion, as described in sections \ref{sec:time-scale} and \ref{sec:data-curation}.  

With upcoming larger datasets containing both more nearby galaxies as well as a larger time baseline of detections -- such as those proposed by Gaia DR4, the Nancy Grace Roman Space Telescope, and the Vera C. Rubin Observatory -- the extragalactic proper motion signal will only become easier to detect \citep{Croft2021}.





\section{Acknowledgments}
    This work has made use of data from the European Space Agency (ESA) mission
    {\it Gaia} (\url{https://www.cosmos.esa.int/gaia}), processed by the {\it Gaia}
    Data Processing and Analysis Consortium (DPAC,
    \url{https://www.cosmos.esa.int/web/gaia/dpac/consortium}). Funding for the DPAC
    has been provided by national institutions, in particular the institutions
    participating in the {\it Gaia} Multilateral Agreement.

    Special thanks to Dr. Coryn Bailer-Jones for his contributions to helping us understand the Gaia time of observations and \texttt{gaiadr3.commanded\_scan\_law} auxiliary data set.

    TZ and KM are supported by Schmidt Sciences. SM is supported by the Engagement of University of Pittsburgh Undergraduates in NASA Science Directorate Programs. RC was supported by NASA grant HST-AR-17554.001-A.
    
\bibliography{main}{}
\bibliographystyle{aasjournal}
\end{document}